\documentclass[11pt]{article}
\usepackage{graphicx}
\usepackage{epsfig}
\usepackage{caption}
\title{\vspace{4cm}\bf 6-arm blue grand design of NGC 309}
\author{A.D.~Chernin\\
Sternberg Astronomical Institute, Moscow University, Moscow,
119991, Russia }

\date{~}

\begin{document}

\maketitle

\begin{abstract}
\noindent {\bf The geometry and physics of the spiral structure of
the giant Hubble type Sc galaxy NGC 309 is studied. A schematic of
two patterns with three arms in each is suggested for the blue
spiral. The red and blue patterns form together a grand design
with two-fold symmetry. A possible gas-dynamics explanation of the
phenomenon is suggested which shows how the two-arm red spiral may
induce the formation of the six-arm coherent blue spiral.}

\vspace{0.3cm}


\end{abstract}

%\vfill \eject

\section{Introduction}

The NGC 309 galaxy is one of the largest and most luminous
grand-design known spirals (Fig.1). At a redshift distance of 83
Mpc (H=70 km/s/Mpc), its diameter is $\simeq$ 70 kpc and its
absolute blue magnitude is -22.52 which corresponds to the
luminosity classification type I (van den Bergh 1960). In the
images of the Hubble Atlas (Sandage 1961) and the Carnegie Atlas
of Galaxies (Sandage and Bedke 1999), the spiral arms of NGC 309
spread over the whole galaxy disk. According to Sandage (1961),
its Hubble type is Sc. Elmegreen and Elmegreen (1984) classify the
galaxy as "extreme grand design".

%Radically different morphologies are observed in optics and
%infrared wavelengths of NGC 309: instead of long, narrow, and
%clumpy multiple arm appearance seen in blue, only two principal
%arms and a prominent central bar are seen in red; the red arms are
%short, wide and smooth (Block and Wainscoat 1991). A similar
%difference between blue and red images was earlier found in the
%M51 galaxy by Zwicky (1957) who concluded that the morphology of
%the old red population of the disk need not mimic the morphology
%of the young blue population. The same point was also made by
%Vorontsov-Vel'yaminov (1978).

%------------------------ Fig. 1
\begin{figure}[bh]
\noindent\centering
{\includegraphics[height=66mm] %\includegraphics[width=83mm,height=66mm]
{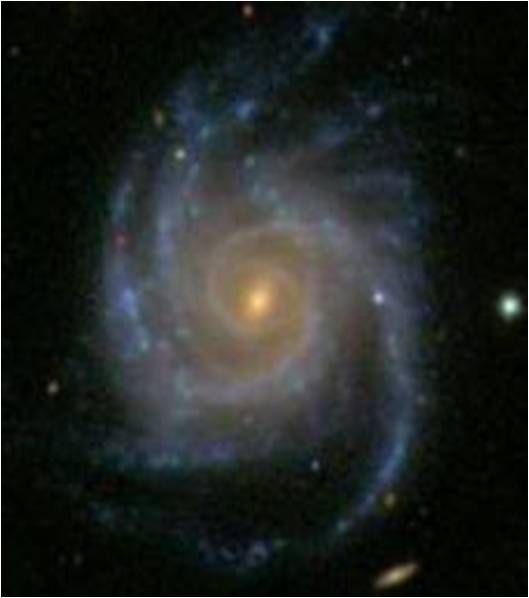} %{CherninFig1.eps}
} \caption{ SDSS image of the spiral galaxy NGC 309.  From SDSS
survey, compose $gri$ image. Attribution for these images ought to
be, at least, David W. Hogg, Michael R. Blanton, and the Sloan
Digital Sky Survey Collaboration.} \label{fig1}
\end{figure}

Radically different morphologies are observed in optics and
infrared wavelengths of NGC 309: instead of long, narrow, and
clumpy multiple arm appearance seen in blue, only two principal
arms and a prominent central bar are seen in red; the red arms are
short, wide and smooth (Block and Wainscoat 1991). A similar
difference between blue and red images was earlier found in the
M51 galaxy by Zwicky (1957) who concluded that the morphology of
the old red population of the disk need not mimic the morphology
of the young blue population. The same point was also made by
Vorontsov-Vel'yaminov (1978).

%------------------------ Fig. 1
%\begin{figure}
%\noindent\centering{
%\includegraphics[width=83mm,height=66mm]{CherninFig1.eps}
%} \caption{ Local flow of giants: Velocity-distance diagram. The
%thick lines are for the basic model. The thin lines are for the
%model without the clusters of galaxies. Solid line is the
%asymptotic linear relation $V = H_{\Lambda} R$. Black dots are for
%the present-day values.} \label{fig1}
%\end{figure}

The morphology features of NGC 309 combined with an almost face-on
orientation offers unique prospects for addressing important
issues associated with the geometrical structure and physical
nature of grand-design spirals. It is also important that the
galaxy do not have close companions; because of this its special
features must completely be due to the internal causes. What is
the physical relation between the two-arm red spiral and the
multi-arm blue spiral of NGC 309? An answer to this question
proposed in this paper is based on the gas-dynamics approach to
the spiral structure formation proposed by Roberts (1969) and
Pikel'ner (1970). It is argued below that the blue multi-arm
grand-design appears as a nonlinear response of the gaseous disk
to the gravitational potential produced by the red arms.

In Sec.2, a schematic for the blue and red spirals of NGC 309 is
suggested based mainly on the images published by Block and
Wainscoat (1991); in Sec.3, the nonlinear gas-dynamics structure,
the "Landau-Lifshitz configuration", is discussed as a possible
physical mechanism of the blue arm formation; the results are
summarized in Sec.5.

\section{Blue and red spirals}

The optical image of NGC 309 presented by Block and Wainscoat
(1991) was processed in a special way, so that faint features were
printed at the same contrast as the bright ones. The processing
enables to recognize six long arms of different brightness, not
saying about small-scale spurs and branches, in the improved blue
image inside 35 kpc from the disk center ($H = 70$ km/s/Mpc).
Block and Wainscoat (1991) mentioned a "three-arm appearance" of
NGC 309 in blue light, and the three of the arms are the brightest
ones indeed; but three others can also be recognized in the image.
In Fig. 2, the six-arm blue spiral is sketched in a manner of
Zwicky's (1959) drawing for M51. The blue grand design of NGC 309
is showed cleaned of small-scale details and irregularities like
spurs, bifurcations, branches, clumps, etc. The lines which
represent the arms are given regardless the brightness or the
width of the arms, in Fig.2.

%------------------------ Fig. 2
%\begin{figure}
%\noindent\centering{
%\includegraphics[width=83mm,height=66mm]
%{CherninFig2.eps} }
%\caption{ A schematic of the six-arm blue
%spiral of the galaxy NGC 309: two coherent patterns with three
%arms in each. The two-arm red spiral is indicated by closed
%contour in the inner area of the disk.} \label{fig2}
%\end{figure}

\begin{figure}[th]
\noindent\centering{
\includegraphics[height=66mm]
{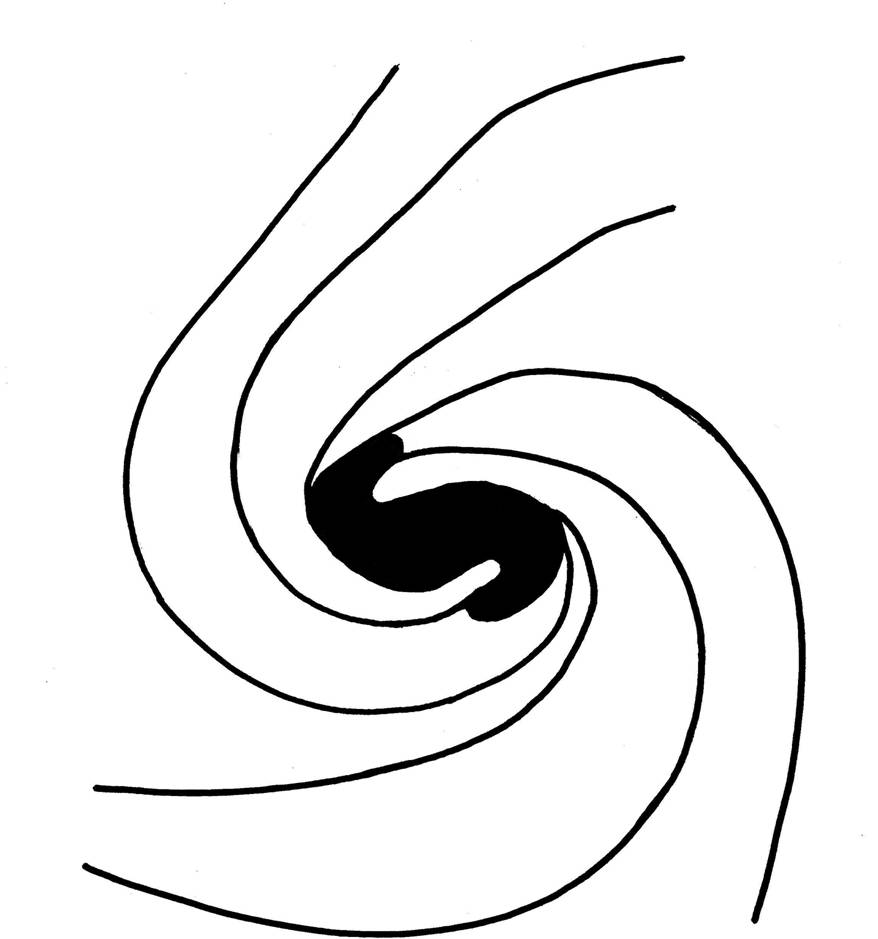} }
\caption{ A schematic of the six-arm blue
spiral of the galaxy NGC 309: two coherent patterns with three
arms in each. The two-arm red spiral is indicated by closed
contour in the inner area of the disk.} \label{fig2}
\end{figure}

Ignoring small-scale features as an image noise, we find the six
blue arms to be well regular and similar in shape to each other,
in Fig. 2. It seems to be especially important that the blue arms
are seen to be arranged into two wide coherent patterns with three
arms in each. These patterns are not completely identical, but
roughly similar in their shape. Each of the patterns goes from the
central region outward, one to the NW direction (North to the top,
East to the left in Fig. 2), and the other goes to SE direction.
The length, width (which increases from the center outwards) and
winding angle are practically the same in both patterns. The two
patterns are located relative to each other in the way that their
phase angles differ from each other in the angle of $\simeq
180^o$. The patterns together form, therefore, a regular
grand-design structure with two-fold symmetry.

The red (near-infrared, in fact) image of NGC 309 (Block and
Wainscoat 1991) shows "only two short principal arms". The red
spiral is of a "two-arm grand-design, with a prominent bar" which
is "recognizable easily and unambiguously". It is characterized by
55-75\% surface brightness enhancement compared with the
underlying disk. The red arms lie within a radius of 17 kpc from
the galaxy center. They are smooth and round and "bear a striking
resemblance to the red arms discovered by Zwicky on composite
photographs of M51". Whereas the winding angles of the blue arms
are large, the surface brightness of the red arms "decreases
abruptly after a winding angle of only $\sim 180^o$". The inner
two-arm structure is quite regular and has obvious two-fold
symmetry, as represented in Fig. 2. Thus, both red morphology and
blue morphology of the galaxy prove to have the same type of
global symmetry. Moreover, the phases of the both are nearly
equal. The same symmetry and the same phase give a clear evidence
for a strong dynamical relation between the red and blue spirals
of NGC 309.

It is hardly possible to recognize with certainty how the blue
patterns are "connected" with the two-arm red spiral in the inner
area of the galaxy where the blue patterns start winding. A
possible version of the connection is given in Fig.2 where we use
essentially a guidance from the Zwicky's (1959) scheme for M51 and
assume that each of the blue patterns starts to develop within the
corresponding red arm.

Note that multi-arm spiral structures may also be found in the
galaxies NGC 753, NGC 2997, NGC 5247, NGC 5584, NGC 5494, NGC
1566, NGC 3423, NGC 6878, NGC 6946, NGC 7125, NGC 7689, etc.

\section{Landau-Lifshitz configuration}

A possible approach to the physics which is behind the visible
geometry of the NGC 309 can be developed on the basis of the
classic theory of spiral shocks (Lin and Shu 1964, Roberts 1969,
Pikel'ner 1970, Shu {\it et al.} 1973, Liverts and Mond 2003).
According to the Roberts-Pikel'ner's spiral-shock paradigm,
large-scale layers of compressed matter form in the interstellar
gas when the gas flow crosses the gravitational potential wells of
the spiral arms. The gravitational potential of the arms is
considered to be mainly due to the density wave in the
distribution of the old populations of red stars in the disk.
Block and Wainscoat (1991) point out that the red image of NGC 309
enables to see the underlying gravitational potential in the disk
of the galaxy. Indeed, the potential has the two-arm structure and
two-fold symmetry, as indicated by the red morphology of the
galaxy (Fig.1).

How can the two-arm gravitational potential drive six-arm grand
design in gaseous disk? This is the key question of the NGC 309
physics.

In a search for an answer to the question, a nonlinear supersonic
gas-dynamical structure may be addressed which was described by
Landau and Lifshitz (1979) as a "non-evolutionary jump". The jump
appears, if, for instance, two gaseous masses come into contact
collision with supersonic velocities. It forms on the surface of
the first contact of the masses. Such a jump is called
non-evolutionary because it is not a result of gradual nonlinear
evolution of a supersonic flow, but produced by two independent
flow masses that did not "know" about each other before the
collision. In a simplest way, one may consider two plane flows
that move towards each other from opposite directions. Generally,
the physical parameters of one of the flows are completely
independent from that of the other. Because of this, there is no
"standard" correspondence between the densities, velocities,
pressure, etc. on the opposite sides of the jump surface, and so
the gas-dynamics laws of continuity are not met at this surface.
The non-evolutionary jump is completely unstable, and it must
rapidly transform into a structure in which the continuity laws
are satisfied (Landau and Lifshitz 1979). As a result, the
non-evolutionary jump transforms into a set of "standard" jumps.
The new structure -- the Landau-Lifshitz (LL) configuration --
includes three basic elements: a contact jump at the surface of
the first contact of the flows and two shock fronts on the
opposite sides of the contact jump. The normal (to the plane of
the contact) component of the flow velocity is continuous at the
contact jump. The flow densities are, generally, different on the
sides of its plane. The matter inflows into the area between the
shock fronts from the opposite sides, and the normal component of
the matter flows has a jump at the surfaces of the chocks. This
configuration is hydrodynamically stable, and a quasi steady-state
can develop at which the configuration may evolve only on the time
scale that is much larger than the time scale of the instability
that leads to formation of the configuration itself. All the three
jumps of the LL-configuration are the surfaces (actually layers)
where gas is compressed.

If the flow is not one-dimensional, an additional dynamical
feature may appear in the LL-configuration: the tangential (to the
contact jump) velocity component may have a jump (tangential
discontinuity) at the contact jump surface. This kind of
discontinuity is known to be unstable, and a turbulence layer
develops usually along the contact jump.

It may be assumed that the LL-configuration can appear in the
gaseous disks of galaxies as the gas-dynamics response to the
gravitational potential produced by over-densities in the red
stellar population of the disk. In NGC 309, the massive two-arm
over-density in the disk central area produces a two-arm spiral
gravitational potential well which extends over the whole rotating
disk (Lin and Shu 1964, Roberts 1969, Pikel'ner 1970, Shu {\it et
al.} 1973, Liverts and Mond 2003). Moving in this potential, the
gas of the rotating disk crosses the potential well and forms the
LL-configuration along each of the two arms of the gravitational
potential. The compression of the gas in the three jumps of the
configuration leads to star formation there. As a result, the
jumps become loci of star formation, and the sites of young blue
stars trace the underlaying gas-dynamics configuration. In this
way, two-arm red spiral can induce six-arm blue spiral in the
galaxy disk.

\section{Conclusions}

The image analysis of Sec.2 suggests a rather regular model for
the spiral structure of the galaxy NGC 309. This model is
represented schematically by Fig. 2. As one can see:

1. The blue spiral grand design has six arms which are similar to
each other in shape, if they are considered regardless their
brightness.

2. The blue arms are arranged into two coherent patterns with
three arms in each.

3. The difference in phase angle between the two blue patterns is
$\simeq 180^o$, and so they form together grand design of two-fold
symmetry. The red spiral has the same symmetry.

The physical interpretation of the phenomenon of the six-arm grand
design in NGC 309 may be searched in the context of the
Roberts-Pikel'ner's spiral shock and the Landau-Lifshitz
configuration. On this basis, a conjecture is suggested according
to which:

1. The two-pattern blue grand design with two-fold symmetry is
driven by the gravitational potential produced by the inner
two-arm red spiral.

2. Each of the blue patterns consists of three arms formed by the
Landau-Lifshitz nonlinear gas-dynamics configuration of the three
standard jumps.

3. Gas compression in the standard jumps triggers star formation
which reveals to us the six-arm underlying nonlinear gas-dynamics
structure.

4. First attempts to find the Landau-Lifshitz configuration in
computer simulations of the spiral structure seem to be promising
(Filistov 2013).

\vspace{1cm}

Discussions with V. Arkhipova, Yu. Efremov, E. Filistov and A.
Zasov are highly appreciated.

%\newpage

{\bf REFERENCES}
\vspace{0.3cm}

Block D.L., Wainscoat R.J. 1991, Nature 353, 48

Elmegreen B.G. and Elmegreen D.M. 1984, ApJ Suppl. 54, 127

Filistov E.A. 2014, Ast.Rep. 58, 451

Greenawalt B., Walterlos R.A.M., Thilker D., Hoopes C.G. 1998, ApJ
506, 135

Hill J.K., Waller W.H., Cornett R.H., {\it et al.} 1997, ApJ 477, 673

Landau L.D. and Lifshitz E.M., 1979, Fluid Mechanics, Pergamon Press,
Oxford

Lin C.C. and Shu F. 1964, ApJ 140, 646

Liverts E. and Mond M. 2003, {\em Physics and Astronomy Galaxies
and Chaos}, Lecture Notes in Physics, 626, 109, Springer, Berlin

Pikel'ner S.B. 1970, Sov. Astron. 47, 254

Roberts W.W. 1969, Ap.J 158, 123

Sandage A. 1961, {\em The Hubble Attlas of Dalaxies}, Carnegie Inst.,
Washington, DC

Sandage A. and Bedke J. 1994, {\em The Carnegie Attlas of Galaxies},
Carnegie Inst., Washington, DC

Shu F.H., Milione V., and Roberts W.W. 1973, ApJ 183, 819

van den Bergh S. 1960, ApJ 131, 215

Vorontsov-Vel'yaminov B.A. 1978, {\em Extragalactic Astronomy}, Nauka,
Moscow (English Edition by Harwood Academic Pubs., New York, 1987)

%Waller W.H., Bohlin R.C., Cornett R.H. {\it et al.} 1997, ApJ 481, 169

%Zwicky F. 1957, {\em Morphological Astronomy}, Springer, Berlin

Zwicky F. 1959, in {\em Handbuch der Physik}, ed. S. Fl\"ugge, Springer,
Berlin

\vspace{1cm}
%\newpage

\end{document}